\documentclass[aps,pra,twocolumn,showpacs,groupedaddress,floatfix]{revtex4}
\usepackage{graphicx}
\usepackage{amssymb,amsmath}
\usepackage{ulem}

\begin{document}

\title{Vortex Dynamics in an Annular Bose-Einstein Condensate}
\author{S. J. Woo} \email{sjwoo@kias.re.kr}
\author{Young-Woo Son} \email{hand@kias.re.kr}
\affiliation{Korea Institute for Advanced Study, Seoul 130-722, Korea}

\date{\today}
\begin{abstract}
We theoretically show that the topology of a non-simply-connected annular 
atomic Bose-Einstein condensate enforces 
the inner surface waves to be always excited with outer surface excitations
and that the inner surface modes are associated with induced vortex dipoles 
unlike the surface waves of a simply-connected one with vortex monopoles.
Consequently, under stirring to drive an inner surface wave, 
a peculiar population oscillation between the inner and outer surface is generated
regardless of annulus thickness.
Moreover, a new vortex nucleation process by stirring is observed
that can merge the inner vortex dipoles and outer vortex 
into a single vortex inside the annulus.
The energy spectrum for a rotating annular condensate
with a vortex at the center also reveals 
the distinct connection of the Tkachenko modes 
of a vortex lattice to its inner surface excitations. 
\end{abstract}


\maketitle
An atomic Bose-Einstein condensate (BEC) has been considered 
as an excellent many-body macroscopic quantum system 
with experimentally controllable parameters such as number density, 
inter-particle interaction, spin, confining potential and so on. 
Recently, topology of BEC also becomes to be important 
as in many different fields of physics. 
For example, the realization of a circular waveguide 
for an atomic BEC ~\cite{Sauer,Gupta} was followed 
by the achievement of a topologically non-simply-connected 
atomic condensate in a magnetic toroidal trap with a plug laser. 
It shows the first observation of atomic superfluid persistent current \cite{Ryu}
which was further refined to stay longer 
than 40 seconds in an all optical trap \cite{Ramanathan}. 

In torus BECs, there have been many interesting phenomena 
related with topological aspect: 
vortex lattice formation \cite{Penckwitt,Aftalion}, 
dynamics of vortices and solitons 
under stirring \cite{Kanamoto,Piazza,Citro, Mason}, 
and dynamics of vortex dipoles \cite{Mason, Martikainen, Neely}. 
In order to understand the formation and decay of vortices in a superfluid 
related with the toroidal persistent current, 
the knowledge of low-lying energy spectrum and dynamics is essential. 
When the torous BEC 
has flat shape with strong confinement along one direction, 
a toroidal condensate has two distinguishable surfaces, inner 
and outer surfaces (called as an annular BEC hereafter).
The normal mode spectrum calculations for such an annular 
BEC available currently, however, have been done only within quasi-1D regime 
where the azimuthal degree of freedom is meaningful \cite{Jackson}, 
while in the experiments radial degree of freedom cannot be ignored \cite{Ryu, Ramanathan}. 

In this Letter, we calculate numerically the normal mode dynamics 
of an annular BEC within mean field regime considering 
the azimuthal and radial degrees of freedom fully 
compatible with current experiments \cite{Ryu, Ramanathan}.
We have found that the excitations 
on the inner surface of a non-rotating annular BEC 
are associated with vortex dipoles rather than monopoles
and that such inner surface excitations 
cannot be excited by themselves. 
Those aspects suggest topologically
distinct phenomena of annular BECs when they are stirred, e.g., 
population oscillations between inner and outer surfaces
and topological excitations by vortex dipoles.
A rotating annular BEC with nontrivial angular momentum 
is achieved by imposing quantized vorticity 
that penetrates the hollow region of the annulus. 
It is also found that such penetrating vortices, 
which would give vortex lattice excitations, Tkachenko modes, 
if they were in a simply-connected BEC, 
play a crucial role in determining the inner surface excitations 
of a rotating annular condensate. 

\begin{figure}[b]
  \includegraphics[width=1.0\columnwidth]{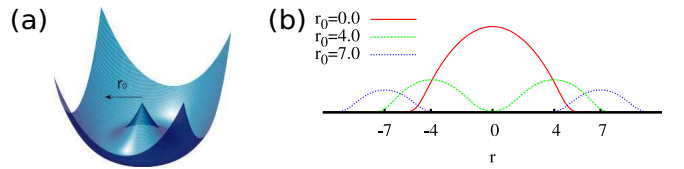}
  \caption{(Color online) 
	 (a) Torus trapping potential. 
	 (b) The radial density distribution of BEC for various torus radii.
  }
  \label{gdstate}
\end{figure}

To describe the normal mode excitations, 
we solve Bogoliubov-de Gennes equations~\cite{Griffin,Leggett} numerically using finite element technique with Hermite polynomial basis~\cite{Baksmaty}; 
${\mathcal L}u_j-g\psi_0^2   v_j=\hbar\omega_ju_j$
and ${\mathcal L}v_j-g\psi_0^{*2}u_j=-\hbar\omega_jv_j$
where ${\mathcal L}=-\frac{\hbar^2\nabla^2}{2M}+V_{\rm{tr}}({\bf r})+2g|\psi_0|^2-\mu$. 
The toroidal trap potential is 
$V_{\rm{tr}}=\frac{M}{2}\left(\omega_{xy}^2(r-r_0)^2+\omega_z^2z^2\right)$ 
where $r=(x^2+y^2)^{1/2}$ as shown in Fig. 1(a). 
Interparticle coupling constant is $g=\frac{4\pi a\hbar^2}{M}$ 
with the $s$-wave scattering length $a$, 
$\mu$ the chemical potential and 
$\psi_0$ the ground state wave function of the condensate. 
$u_j$, $v_j$, and $\omega_j$ represent the wave functions and 
eigenenergy for the $j$-th quasiparticle excitation, respectively. 
$\psi_0$ and $\mu$ are determined 
by solving time-independent Gross-Pitaevskii equation (GPE), 
$\left(-\frac{\hbar^2\nabla^2}{2M}+V_{\rm{tr}}({\bf r})+g|\psi_0|^2-\mu\right)\psi_0=0$
where $\psi_0$ is normalized by the total number of particles, 
$N={\int}d{\bf r}~ |\psi_0|^2$. 
Strong confinement along the $z$-axis, $\omega_{xy} \ll \omega_z$, 
is assumed.
All values here are based on the dimension of the trap potential~\cite{trap}.
Fig.~\ref{gdstate} (b) is the density of condensate 
along the radial direction for 
$N=5000$ and $g=0.1$ for different $r_0$ values. 

Normal mode dynamics can be visualized by noting 
that $\psi_j(t) = \psi_0 + \sqrt{n_q}\left(u_je^{-i\omega_j} - v_j^*e^{i\omega_j}\right)$ 
satisfies time-dependent GPE \cite{Griffin,Leggett}. 
Fig. 2 shows the energy {\it vs.} angular momentum dispersion relations 
of a BEC for different $r_0$ values: (a) $r_0 =0$ and (b) $r_0 =2$ 
correspond to simply-connected BECs, 
(c) $r_0 =4$ for the transition region, and (d) and (e) ($r_0 >4$) 
are for annulus cases. 
We categorize the normal modes using a pair of  
the radial ($n$) and angular ($m$) quantum numbers, i.e., $(n,m)$ mode, 
as in a simply-connected case; $n$ and $m$ represent 
the number of nodes along the radial and angular direction respectively \cite{Woo}. 
Here we notice that the energy dispersion curves as functions of $m$ 
for either simply connected or annular condensate 
show well separated classes of normal modes with quantum number $n$'s. 

\begin{figure}[t]
  \includegraphics[width=1.0\columnwidth]{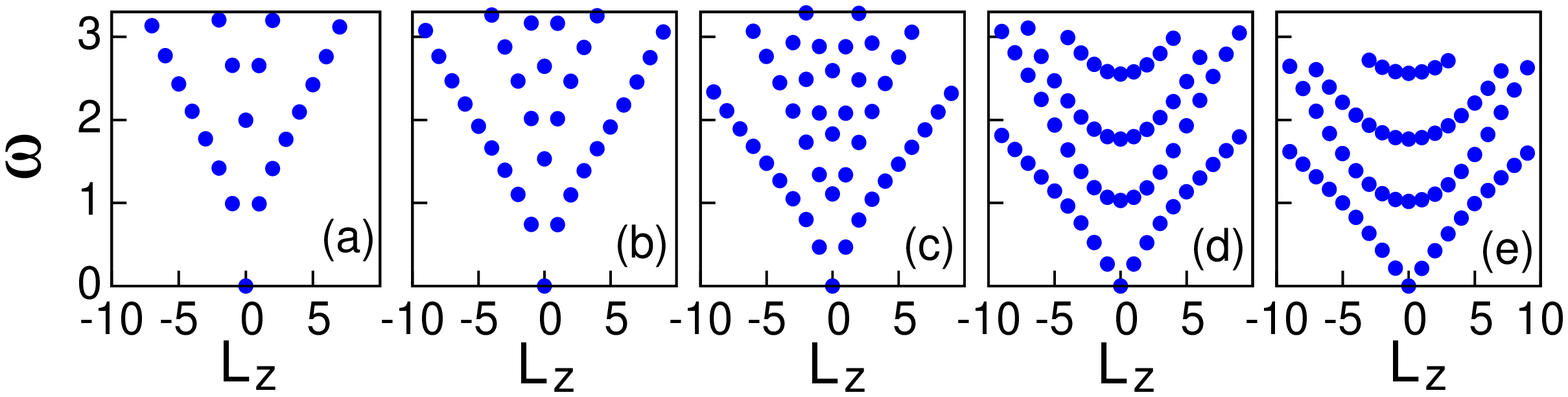}
  \caption{(Color online)  Angular momentum ($L_z$) vs. energy ($\omega$) curves
	  for different $r_0$ values for a non-rotating condensate: $r_0=$ 0 (a), 2 (b), 4 (c), 6 (d) and 7 (e).
}
  \label{LzvsE}
\end{figure}

\begin{figure}[t]
 \includegraphics[width=1.0\columnwidth]{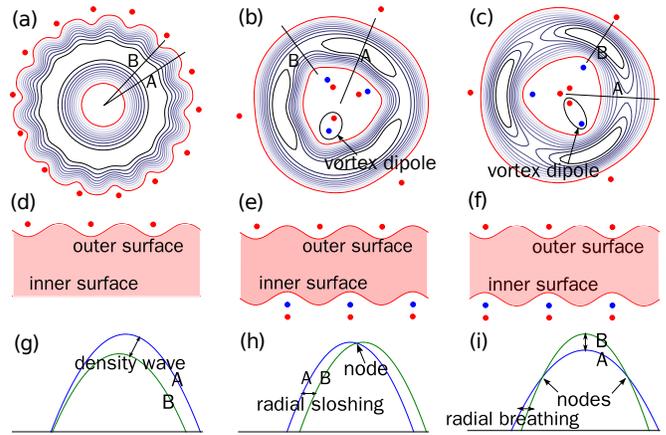}
 \caption{(Color online) Contours for snapshots of $|\psi_j(t)|^2$  for modes
	 with $(n, m)=$ (a) $(0,14)$ (b) $(1,3)$ and (c) $(2,3)$. 
     Darker blue contour lines represent higher densities 
     while red lines are drawn to clarify the density edges. 
	 Red (blue) dots represent induced (anti)vortices associated with the excited normal modes
	 propagating conterclockwise and clarify where the wave nodes are. 
     Contours (a), (b), and (c) are schematically unwound 
	 in (d), (e) and (f) with vortex singularities propagting leftwards, 
	 respectively. Cross sectional density fluctuations 
	 along the lines A and B in (a), (b), and (c)
     are schematically drawn in (g), (h) and (i), respectively.}
  \label{normal modes}
\end{figure}

First, we have found that the surface waves 
with the lowest radial quantum number ($n=0$) 
cannot be generated along the inner surface of an annulus 
even though an additional surface forms on its inner side. 
As $r_0$ changes from 0 to 7 crossing the critical point at $r_0\approx 4$, 
every normal mode can adiabatically be transformed.  
The modes in the lowest curve ($n=0$) in Figs.~\ref{LzvsE}~(d) or (e) 
correspond to normal modes with longitudinal density wave 
propagating along the radial direction. 
Especially with higher $|m|$, $n=0$ density waves appear mainly 
as surface waves mostly on the outer surface as in the simply-connected case \cite{Woo}; 
Fig.~\ref{normal modes}~(a) shows the contour plot of a snapshot 
for the density of $(0,14)$ normal mode.
Fig.~\ref{normal modes}~(d) represents the inner and outer edges 
if the annulus were to be unwound with the corresponding induced vortex singularity positions. 
Another schematic plot, Fig.~\ref{normal modes}~(g), shows two density plots 
along the radial direction at two different angular positions $A$ and $B$ 
in Fig.~\ref{normal modes}~(a) 
which shows the density fluctuation at a certain fixed angular position.

Next, we show that, unlike the case for the lowest radial mode ($n=0$), 
both inner and outer surfaces host surface wave excitations for higher radial modes. 
In Fig.~\ref{LzvsE}~(d) or (e), the next lowest curve ($n=1$) 
corresponds to out-of-phase surface excitations with both inner 
and outer surface waves,
of which wave nodes are placed alternatively along the angular direction. 
The corresponding modes in a simply-connected case, 
Fig.~\ref{LzvsE}~(a) or (b), are those with one radial node 
including breathing mode~\cite{Woo}. 
The $(1,0)$ breathing mode in a simply-connected case has 
the well-known many-body breathing frequency $2\omega_{xy}$ \cite{Pitaevskii}.
One can see that such a breathing mode would transform 
into a radial sloshing mode centered at $r=r_0$
as the condensate gets annular-shaped and 
hence the frequency becomes corresponding radial trap frequency, $\omega_{xy}$. 

Figure~\ref{normal modes}~(b) shows the snapshot of $(1,3)$ mode. 
It should be emphasized that the inner surface wave 
is accompanied by vortex-antivortex pairs, {\it vortex dipoles}, 
differently from the outer surface wave 
that is accompanied by induced vortices or antivortices 
depending on the direction of the wave. 
The critical difference between the two surface waves 
is that the associated vortex singularities 
for an inner surface wave are enclosed by the annulus of condensate 
while those for an outer one are not. 
Due to this, the induced vorticities on the inner surface will 
inevitably generate supercurrent, a topological excitation, 
while those on the outer surface do not. 
By introducing a vortex dipole inside the annular BEC
the vortex or antivortex whichever is closer to the inner surface 
makes a propagating node in the surface wave 
while the other can annihilate the generated supercurrent. 
This property, which is a key finding in this work, 
is attributed to the topological aspect of annular geometry 
and the vortex excitations. 
A schematic plot, Fig.~\ref{normal modes}~(e), shows that 
an out-of-phase surface wave is in fact a transverse density wave. 
Furthermore, Fig.~\ref{normal modes}~(h) clarifies 
that the number of radial nodes ($n$) 
should be one for this specific radial center-of-mass dynamics, 
i.e., radial sloshing. 

The third curve ($n=2$) in Fig.~\ref{LzvsE}~(d) or (e) 
corresponds to in-phase excitations with both inner 
and outer surface waves,
of which wave nodes are placed at the same angular positions. 
The corresponding modes in a simply-connected case, Fig.~\ref{LzvsE}~(a), 
are modes with two radial nodes \cite{Woo}. 
As $r_0$ increases, modes with $n=2$ make radial breathing dynamics 
centered at $r=r_0$ so that the frequency converges to near $2\omega_{xy}$ for $m=0$. 
The $(2,3)$ mode [Fig.~\ref{normal modes}~(c)] 
also shows vortex dipoles for the inner surface excitation. 
Figure~\ref{normal modes}~(f) is a schematic plot showing more clearly 
that an in-phase surface wave is a longitudinal density wave 
while Fig.~\ref{normal modes}~(i) shows that 
the number of radial nodes is two. 
In general, modes with higher $n$ show similar surface waves, 
in-phase for even $n$'s and out-of-phase for odd $n$'s 
while they have additional radial nodes. 

\begin{figure}[b]
 \includegraphics[width=1.0\columnwidth]{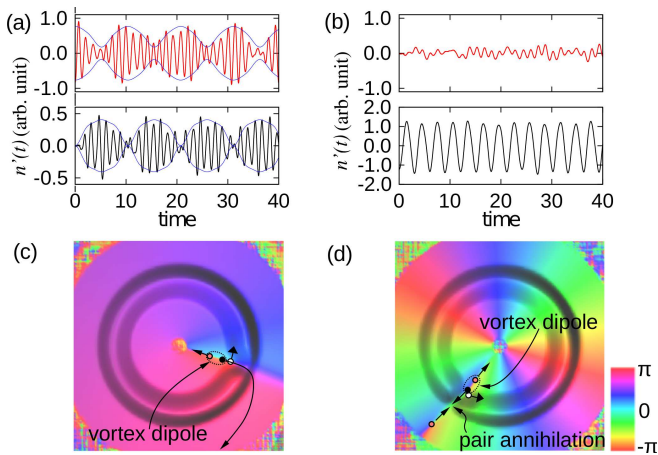}
  \caption{(Color online) The density fluctuation 
	  ($n'(t)=\psi(t)^2-\psi_0^2$~\cite{Woo}) 
	  of the inner (red) and outer (black) surface at $\theta=0$ as a function of time 
	  when either only (a) the inner or (b) outer is excited at $t=0$. 
      Thin blue lines in (a) are guides for eyes.
      Snapshot of $\psi(t)=\sqrt{\rho}e^{i\phi}$ under gaussian stirrer (white dot)
	  with slow (c) and fast (d) speed.
      The height of the surface represents $\sqrt{\rho}$ while the color $\phi$. 
      Red (black) circles represent (anti)vortices
      around which $\phi$ has $\pm 2\pi$ winding. The black arrows in (c) and (d) denote
	  the motions of vortices.
  }
  \label{population oscillation}
\end{figure}

Having categorized all the normal modes, 
it is noted that an inner surface wave should always be 
excited with an outer surface wave, either in-phase or out-of-phase. 
This implies that,
if one drives an inner surface wave by stirring, 
it will not stay steady. 
It is rather a combination of different in-phase and out-of-phase states 
which will generate beating, hence showing population oscillation 
with outer-surface wave independent of the annulus thickness.
In order to confirm this, we use a modified trap potential 
mimicking inner surface mode such as
$V_\epsilon = V_{\rm {tr}}\left(1+\epsilon \cos(k\theta)\right)$ when $r<r_0$ and
otherwise $V_\epsilon = V_{\rm {tr}}$
($k$ is a non-zero integer). 
We obtain the ground state $\psi_\epsilon$ for $V_\epsilon$ 
that has an inner surface density modulation.
Then let it evolve in time in $V_{\rm {tr}}$ after turning off $\epsilon$ to be zero. 
Our simulation for an annular BEC of $r_0=7$
shows that the inner surface density modulation propagates
through the annulus reaching the outer surface and then 
back onto the inner, 
making a population oscillation between the two surface waves
[Fig.~\ref{population oscillation}~(a)].
On the other hand, we have also confirmed that an initial density modulation 
on the outer surface makes a density fluctuation only there without
any penetration at all [Fig.~\ref{population oscillation}~(b)].

Another peculiar topological feature is 
in the nucleation process of vortices that
can introduce a net topological winding number within the annulus.
As shown in Fig.~\ref{normal modes}(c), 
the annulus has a vortex-antivortex pair inside
while a vortex outside along the same angular position. 
If the annulus is stirred for inner and outer excitations to 
merge, they will annihilate themselves leaving a vortex inside the annulus, 
a topological excitation.
Such vortex nucleation procedure cannot occur in a simply-connected case
where phase singularities are typically generated 
outside and then get into the condensate via surface excitations.
In order to confirm this senario, 
we have simulated stirring an annular BEC ($r_0=10$)
with a narrow gaussian stirrer on the inner surface.
Initially, the ground state is obtained with a gaussian stirrer 
fixed on the inner edge and then the speed of the stirrer is ramped up 
along the inner surface adiabatically 
at the angular acceleration of $10^{-5}\omega^2_{xy}/\pi^2$.
Figure 4 (c) shows the emergence 
of the first vortex dipole inside the annulus 
followed by the decay of the antivortex part of the dipole 
through the density dip near the stirrer 
leaving a vortex within the annulus generating supercurrent.
Figure 4 (d) shows a similar case 
with two vortices already within the annulus at higher stirrer speed; 
a vortex dipole inside the annulus emerges together with a vortex outside 
followed by pair annihilation of the antivortex part of the dipole 
and the vortex outside on the rim of the annulus adding another vortex within the annulus.

\begin{figure}[b]
 \includegraphics[width=1.0\columnwidth]{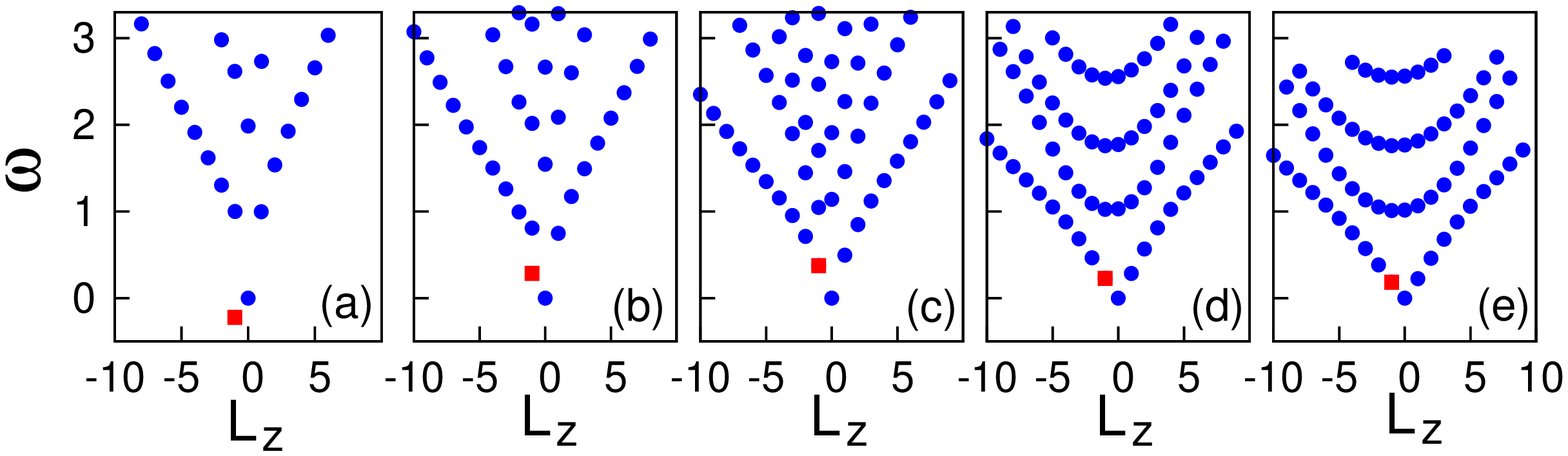}
  \caption{(Color online) $L_z$ {\it vs}. $\omega$
	  for different $r_0$ values for a condensate with a vortex 
	  at the center: $r_0=$ 0 (a), 2 (b), 4 (c), 6 (d) and 7 (e). 
      Red rectangle corresponds to a vortex precession mode at $m=-1$.}
  \label{LzvsEv}
\end{figure}

Now, let us look into the dynamics of an annular BEC with perpetual supercurrent; 
how the low-lying dynamics is affected 
by the current-carrying vortex at the torus center. 
Our calculations have been done in the lab frame. 
Figure~\ref{LzvsEv} is the rotating-case counterpart of Fig.~\ref{LzvsE} 
with a single vortex at the center penetrating the BEC. 
Figure~\ref{LzvsEv}~(a), a simply-connected case, 
is similar to Fig.~\ref{LzvsE}~(a) except that 
it has a negative energy mode at $m=-1$. 
This mode corresponds to the typical vortex precession 
that is the building block for Tkachenko modes 
in the case of vortex lattices~\cite{Fetter}. 
It is interesting to note that Fig.~\ref{LzvsEv}~(e), 
an annulus case when compared to Fig.~\ref{LzvsE}~(e), 
does not show such additional mode even though 
one can still make adiabatic deformation from Fig.~\ref{LzvsEv} (a) to (e). 
We confirm that during the deformation, 
the vortex precession mode with a negative energy 
in Fig.~\ref{LzvsEv} (a) adiabatically shifts to $(0,-1)$ 
mode in Fig.~\ref{LzvsEv}~(e).

This unusual adiabatic stabilization of a vortex precession mode 
is made possible by the changes of the topology. 
As $r_0$ increases deforming the simply-connected condensate 
into an annular one, the negative energy quasiparticle mode that
makes the vortex state energetically unstable as a ground state
acquires positive energy at  $r_0\approx 0.8$ 
because of the vortex pinning by the central potential bump. 
Once density hole is formed in the middle of the BEC 
with a large enough $r_0$ the precessing vortex behaves 
like the induced vortex singularities for an outer surface wave
exciting the corresponding inner surface wave. 
This eventually forms annular density wave yielding 
$(0, -1)$ mode. 

\begin{figure}[t]
 \includegraphics[width=1.0\columnwidth]{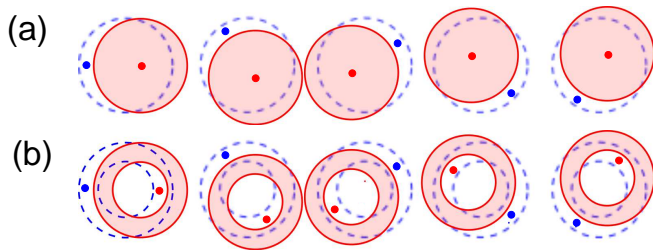}
 \caption{(Color online) (a) Temporally successive (from left to right) 
 snapshots of $(0,-1)$ mode for a rotating simply-connect condensate and 
 (b) those of $(1,-1)$ mode for a annular condensate with perpetual current. 
 Blue dashed circles show the positions of ground state BEC. 
 The red dot for each plot is the vortex that gives perpetual current to the BEC
 while the blue dot represents the induced antivortex accompanied by the $m=-1$
 excitation.}
  \label{vortex to sloshing}
\end{figure}

On the other hand, a mode with $(0,-1)$ in a simply-connected condensate 
shows center-of-mass motion that makes a clockwise circular sloshing dynamics 
around a small circle with the vortex sympathetically following the density maximum
[Fig.~\ref{vortex to sloshing}~(a)]. 
Such a sympathetic motion of the vortex is a resonant behavior 
against the collective hydrodynamic excitation~\cite{Woo}. 
As the condensate transforms into an annular shape, 
this vortex motion also changes into a inner surface wave 
such that the nodes of inner and outer surface wave are out of phase, 
which corresponds to $(1,-1)$ mode [See Fig.~\ref{vortex to sloshing}~(b)]. 
It is important to note that only $m=-1$ modes 
can make resonant dynamics with the singly-quantized vortex 
at the center of a rotating simply-connected BEC.
With such resonant dynamics, $(n,-1)$ mode in a simply-connected BEC
adiabatically deforms into $(n+1,-1)$ mode in an annular BEC.
This resonant behaviour lowers the energy of the normal mode 
so that the $m=-1$ mode has the lowest frequency 
for each family of $n$ in a rotating annular BEC 
with a singly-quantized vortex at the center. 
This can be generalized to
a rotating annular BEC with a vortex 
with higher winding number $m_{\rm v}>1$
such that a mode with $(0,-m_{\rm v})$ 
shows the characteristics of hydrodynamic surface excitations
on the inner surface with $m_{\rm v}$ nodes 
while it can be adiabatically transformed 
from the Tkachenko mode of a vortex lattice in a simply-connected case. 
%Such an intermediate regime between Tkachenko modes and 
%hydrodynamic excitations was also observed in a different context, spinor BEC \cite{Woo2}. 

In summary, our study on the Bogoiubov-de Gennes excitations 
of annular BEC reveals that 
the additional inner surface excitations are analytical continuations 
of collective excitations of a simply-connected condensate 
and that the topological aspect of the annulus forces 
inner surface waves associated with vortex dipoles.
These aspects give rise to interesting topological phenomena such as 
population oscillation and new vortex nucleation process.
It is further found that, with perpetual current, the inner surface excitations
have distinct connection to the the Tkachenko modes in a simply-connected 
condensate.

{\it Note added.} Recently, we learned of a related preprint~\cite{Dubessy}.\\

We thank Prof. Jung Hoon Han for fruitful discussions.
Y.-W.S.was supported by the NRF grant funded 
by the Korea government (MEST) (QMMRC, No. R11-2008-053-01002-0)
and the CAC of KIAS.


\begin{thebibliography}{99}

\bibitem{Sauer} J.~A.~Sauer, M.~D.~Barrett and M.~S.~Chapman, Phys.~Rev.~Lett. {\bf 87}, 270401 (2001). 
\bibitem{Gupta}S.~Gupta {\it et al.}, Phys.~Rev.~Lett. {\bf 95}, 143201 (2005).
\bibitem{Ryu} C.~Ryu {\it et al.}, Phys.~Rev.~Lett. {\bf 99}, 260401 (2007).
\bibitem{Ramanathan} A.~Ramanathan {\it et al.}, Phys. Rev. Lett. {\bf 106}, 130401 (2011).
\bibitem{Penckwitt} A.~A.~Penckwitt, R.~J.~Ballagh and C.~W.~Gardiner, 
	                         Phys.~Rev.~Lett. {\bf 89}, 260402 (2002).
\bibitem{Aftalion}A.~Aftalion and P.~Mason, Phys.~Rev.~A {\bf 81}, 023607 (2010).
\bibitem{Kanamoto} R.~Kanamoto, L.~D.~Carr and M.~Ueda, Phys.~Rev.~Lett. {\bf 100}, 060401 (2008).
\bibitem{Piazza}F.~Piazza, L.~A.~Collins and A.~Smerzi, Phys.~Rev.~A. {\bf 80}, 021601(R) (2009). 
\bibitem{Citro}R. Citro, A.~Minguzzi and F.~W.~J.~Hekking, J.~Phys.: Conf.~Ser. {\bf 150}, 032015 (2009).
\bibitem{Mason} P.~Mason and N.~G.~Berloff, Phys.~Rev.~A. {\bf 79}, 043620 (2009).
\bibitem{Martikainen} J.-P.~Martikainen {\it et al.}, Phys.~Rev.~A. {\bf 64}, 063602 (2001).
\bibitem{Neely} T.~W.~Neely {\it et al.}, Phys.~Rev.~Lett. {\bf 104}, 160401 (2010).
\bibitem{Jackson} A.~D.~Jackson and G.~M.~Kavoulakis, Phys.~Rev.~A {\bf 74}, 065601 (2006).
\bibitem{Griffin} A.~Griffin, Phys.~Rev.~B {\bf 53}, 9341 (1996). 
\bibitem{Leggett}A.~J.~Leggett, Rev.~Mod.~Phys. {\bf 73}, 307 (2001).
\bibitem{trap}	The time, length, energy, and angular momentum are measured 
in the units of $\omega_{xy}^{-1}$, $\sqrt{\hbar/M\omega_{xy}}$, $\hbar \omega_{xy}$ 
and $\hbar$. 
\bibitem{Woo} S.~J.~Woo, L.~O.~Baksmaty, S.~Choi and N.~P.~Bigelow, Phys.~Rev.~Lett. {\bf 92}, 170402 (2004).
\bibitem{Baksmaty} L.~O.~Baksmaty, S.~J.~Woo, S.~Choi and N.~P.~Bigelow, Phys.~Rev.~Lett. {\bf 92}, 160405 (2004).
\bibitem{Pitaevskii} L.~P.~Pitaevskii and A.~Rosch, Phys.~Rev.~A {\bf 55}, R853 (1997).
\bibitem{Fetter} A.~L.~Fetter and A.~A.~Svidzinsky, J.~Phys. {\bf 13}, 135 (2001).
\bibitem{Dubessy} R.~Dubessy, T.~Liennard, P.~Pedri, and H.~Perrin (2012), arXiv:1204.6183v1.
%\bibitem{Woo2} S.~J.~Woo, S. Choi, L.~O.~Baksmaty, and N.~P.~Bigelow, Phys.~Rev.~A {\bf 75}, 031604, (2007).
\end{thebibliography}
\end{document}